\documentclass{elsart3}
\usepackage{graphics}
\usepackage{epsfig}

\usepackage{amssymb}

\begin{document}

\begin{frontmatter}

\title{Scale-free avalanche dynamics in the stock market}

\author{M. Bartolozzi,\thanksref{a}
}
\ead{mbartolo@physics.adelaide.edu.au}
\author{D. B. Leinweber,\thanksref{a}} 
\author{A. W. Thomas\thanksref{a}\thanksref{b}}
\address[a]{Special Research Centre for the Subatomic 
Structure of Matter (CSSM) and Department of Physics, University of
Adelaide, Adelaide, SA 5005, Australia}
\address[b]{Jefferson Laboratory, 12000 Jefferson Ave., Newport News,
  VA 23606, USA}

\begin{abstract}
Self-organized criticality has been claimed to play an important
role in many natural and social systems.  In the present work we
empirically investigate the relevance of this theory to stock-market
dynamics.
Avalanches in stock-market indices are identified using a multi-scale
wavelet-filtering analysis designed to remove Gaussian noise from the
index.  Here new methods are developed to identify the optimal
filtering parameters which maximize the noise removal.
The filtered time series is reconstructed and compared with the
original time series.
A statistical analysis of both high-frequency Nasdaq E-mini Futures
and daily Dow Jones data is performed.
The results of this new analysis confirm earlier results revealing a
robust power law behaviour in the probability distribution function of
the sizes, duration and laminar times between avalanches.  This power
law behavior holds the potential to be established as a stylized fact of
stock market indices in general.  
While the memory process, implied by the power law distribution of the
laminar times, is not consistent with classical 
models for self-organized criticality, we note that a power-law
distribution of the laminar times cannot be used to rule out
self-organized critical behaviour.
\end{abstract}

\end{frontmatter}

\section{Introduction}

Attracted by several analogies with the dynamics of natural systems,
physicists, especially during the last decade, have attempted to
understand the mechanism behind stock-market dynamics by applying
techniques and ideas developed in their respective
fields~\cite{mantegna}.

In this context, possible connections between self-organized
criticality (SOC) and the stock market, or economics in general, have
been investigated
theoretically~\cite{Bak93b,Bak97,Turcotte99,Feigenbaum03,Bartolozzi05}.
The theory of SOC, originally proposed in the late eighty's by Bak,
Tang and Wiesenfeld (BTW)~\cite{Bak8788} to explain the ubiquity of
power laws in Nature, is claimed to be relevant in several different
areas of physics as well as biological and social
sciences~\cite{Turcotte99,Jensen}.  The key concept of SOC is that
complex systems -- {\it i.e.}~systems constituted by many non-linear
interacting elements -- although obeying different microscopic
physics, may {\em naturally} evolve toward a {\em critical} state
where, in analogy with physical systems near the phase transition,
they can be characterized by power laws.  The critical state is an
ensemble of metastable configurations and the system evolves from one
to another via an avalanche-like dynamics~\cite{Jensen}.

The classical example of a system exhibiting SOC behaviour is the 2D
sandpile model~\cite{Bak8788,Jensen,Turcotte99}.  Here the cells of a
grid are randomly filled, by an external driver, with ``sand''.  When
the gradient between two adjacent cells exceeds a certain threshold a
redistribution of the sand occurs, leading to more instabilities and
further redistributions.  The benchmark of this system, indeed of all
systems exhibiting SOC, is that the distribution of the avalanche
sizes, their duration and the energy released, obey power laws.  As
such, they are {\em scale-free}.
 
In the present work we search for imprints of SOC in the stock market
by studying the statistics of the coherent periods (that is, periods
of high volatility), or {\em avalanches}, which characterize its
evolution.  
We analyze the tick-by-tick behaviour of
the Nasdaq E-mini Futures (NQ) index, $P(t)$, from 21/6/1999 to
19/6/2002 for a total of $2^{19}$ data. In particular, we study the
logarithmic returns of this index, which are defined as $r(t)=\ln
\left[ P(t)/ P(t-1) \right]$.  Possible differences between daily and
high frequency data have also been taken into consideration through
the analysis of the Dow Jones daily closures (DJ) from 2/2/1939 to
13/4/2004, for a total of $10^{14}$ data.

This work extends our earlier work on this subject \cite{SOC1} by
introducing new criteria to optimize the filtering of the time series
essential to separating quiescent and avalanche dynamics.  The
properties of the time series reconstructed from the filtered returns
are also examined.  The issue regarding the presence of SOC in the
stock market is of not only of theoretical importance, since it would
lead to improvements in financial modeling, but could also enhance the
predictive power~\cite{Caglioti95} of Econophysics.

In the next section we present the analysis methodology while in
Sec.~\ref{data_analysis} the results of the analysis are presented.
Discussions and conclusions are contained in the last section.

\section{Avalanche Identification via Wavelets} 
\label{sec:wavelet}

The logarithmic returns of stock indices rarely display intervals of
genuinely quiescent periods, yet such periods are vital to the
quantitative identification of avalanche dynamics.  As such, noise must
be filtered from the time series.
Ideally, only Gaussian noise, associated with the {\em efficient}
phases of the market where the movements can be well approximated by a
random walk~\cite{mantegna}, is to be filtered from the time-series
returns.  Such dynamics have no memory and contrast the avalanche
dynamics, {\it i.e.} anomalous periods characterized by large
fluctuations, that we aim to analyze.

Naively, one might simply set a threshold for the logarithmic returns,
below which the index is deemed to be laminar.  However, a simple
threshold method is not appropriate, as it would include in the
filtering some non-Gaussian returns at small scales that are relevant
in our analysis.

\begin{figure}
\centerline{\epsfig{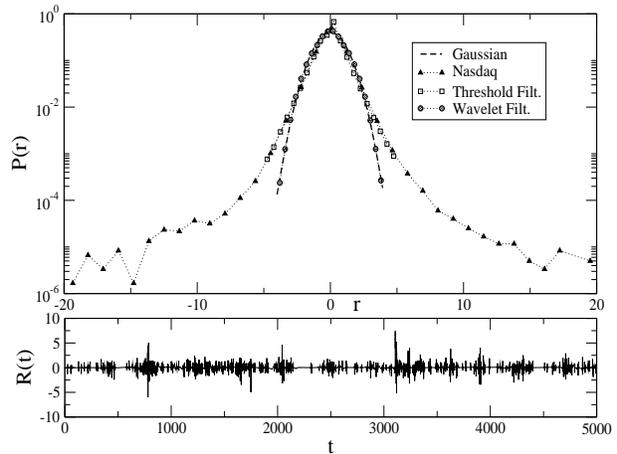}}
\caption{ (Top) Comparison between the PDF of the original time series
for the NQ index and its wavelet filtered version for $C=1$.  A
Gaussian distribution is plotted for visual comparison.  The simple
threshold, $r_{th}=5$, method for filtering is also shown.  In this
case it is clear that we do not remove just Gaussian noise, but also
coherent events that can be relevant for the analysis.
(Bottom) A window of the time series of the residuals obtained by
subtracting filtered time series from the original time series.
Avalanches of high volatility contrast periods of genuinely quiescent
behavior.  All the data in the plots have been standardized, $r(t)
\rightarrow (r(t)-\langle r \rangle)/\sigma(r)$, where $\langle
... \rangle$ and $\sigma(r)$ are, respectively, the average and the
standard deviation during the period under study.}
\label{filt_pl}
\end{figure}

This difficulty is illustrated in Fig.~\ref{filt_pl} (Top) where the
probability distribution function (PDF) for the returns of the NQ
index, filtered using a fixed threshold of $r_{th}= 5$ standard
deviations is shown by the open squares.  In this case broad wings,
related to events that do not follow Gaussian statistics, are clearly
evident.

However, an important {\em stylized fact} of financial returns -- the
{\em intermittency} of financial returns~\cite{mantegna} -- can be
used to identify an appropriate filtering scheme.  Already, physicists
have drawn analogies with the well known phenomenon of intermittency
in the spatial velocity fluctuations of hydrodynamic
flows~\cite{Frisch,Ghashghaie96,Mantegna97}.  Both systems display
broad tails in the probability distribution function~\cite{mantegna}
and a non-linear multifractal spectrum~\cite{Ghashghaie96} as a
result of this feature.
The empirical analogies between turbulence and the stock market
suggest the existence of a temporal information cascade for the
latter~\cite{Ghashghaie96}.  This is equivalent to say that
various traders require different information according to their
specific strategies.  In this way, different time scales become
involved in the trading process.  

In the present work we use a wavelet method in order to study
multi-scale market dynamics.  The wavelet transform is a relatively
new tool for the study of intermittent and multifractal
signals~\cite{Farge92}.  This approach enables one to decompose the
signal in terms of scale and time units and so to separate its
coherent parts -- {\it i.e.} the bursty periods related to the tails
of the PDF -- from the noise-like background.  This enables an
independent study of the avalanches and the quiescent
intervals~\cite{Farge99}.

The wavelet transform (WT) is defined as the scalar product of the
analyzed signal, $f(t)$, at scale $\lambda$ and time $t$, with a real
or complex ``mother wavelet'', $\psi(t)$.  In the discrete wavelet
transform (DWT) case, used herein, this reads:
\begin{eqnarray}
W_{T}f(t) & = &\frac{1}{\sqrt{\lambda}}\int f(u) \,
{\psi}\!\left (\frac{u-t}{\lambda} \right ) \, du \, ,\\ \nonumber 
& = & 2^{-j/2}\int
f(u)\: \psi(2^{-j}u-n)\,  du \, ,
\label{dwt}   
\end{eqnarray}   
where the mother wavelet is scaled using a dyadic set.  One chooses
$\lambda=2^{j}$, for $j=0,...,L-1$, where $\lambda$ is the scale of
the wavelet and $L$ is the number of scales involved, and the temporal
coefficients are separated by multiples of $\lambda$ for each dyadic
scale, $t=n 2^{j}$, with $n$ being the index of the coefficient at the
$j$th scale.

The wavelet coefficients are a measure of the correlation between the
original signal, $f(t)$, and the mother wavelet, $\psi(t)$, at scale
$j$ and time $n$.  In the analysis presented in the next section, we
use the Daubechies--4 wavelet as the orthonormal
basis~\cite{Daubechies88}.  However, tests performed with different
sets do not show any qualitative difference in the results.

The utility of the wavelet transform in the study of turbulent signals
lies in the fact that the large amplitude wavelet coefficients are
related to the extreme events corresponding to the tails of the PDF,
while the laminar or quiescent periods are related to the coefficients
with smaller amplitude~\cite{Kovacs01}.  In this way, it is possible
to define a criterion whereby one can filter the time series of the
coefficients depending on the specific needs.  In our case, we adopt
the method used in Ref.~\cite{Kovacs01} and originally proposed by
Katul et al.~\cite{Katul94}.  In this method wavelet coefficients that
exceed a fixed threshold are set to zero, according to
\begin{equation}
\tilde{W}_{j,n}=\left \{  \begin{array}{ccc} W_{j,n} & {\rm if} &
 W^{2}_{j,n}<C \,\langle W^{2}_{j,n}\rangle_{n} \, .\\
0 & {\rm otherwise} \, .
\end{array} \right.
\end{equation}
Here $\langle \ldots \rangle_{n}$ denotes the average over the time
parameter $n$ at a certain scale $j$ and $C$ is the threshold
coefficient.  In this way only the dynamics associated with the
efficient phases of the market where the movements can be well
approximated by a random walk~\cite{mantegna} are preserved.

Once we have filtered the wavelet coefficients $\tilde{W}_{j,n}$ an
inverse wavelet transform is performed, obtaining what should
approximate Gaussian noise.  The PDF of this filtered time series is
shown, along with the original PDF in Fig.~\ref{filt_pl} (Top).  It is
evident how the distribution of the filtered signal matches perfectly
a Gaussian distribution.

In the same figure (Bottom), we also show the logarithmic returns,
$R(t)$, of the original time series after the filtered time series has
been subtracted.  Truly quiescent periods are now evident,
contrasting the the bursty periods, or avalanches, which we aim to
study.

\begin{figure}
\hspace{-0.3cm}
\centerline{\epsfig{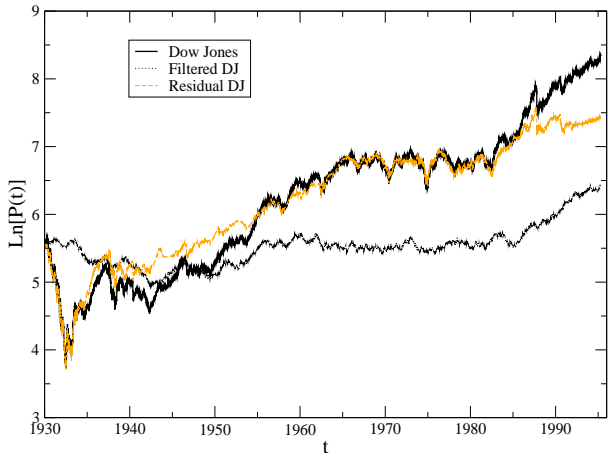}}
\caption{ The Dow Jones time series is superimposed with the time
  series reconstructed from the filtered returns and the residual
  returns remaining after the filtered returns are subtracted from the
  original returns.  The price behaviour generated by the
  ``efficient'' or filtered returns is largely independent of the
  observed price.  The filtering parameter, in this case, is $C=1$.}
\label{ts_pl}
\end{figure}

The time series of logarithmic prices is reconstructed from the residuals in
Fig.~\ref{ts_pl} and is contrasted with the one reconstructed 
from the filtered Gaussianly distributed returns. 
 Note how, in the latter case, 
the time series is completely independent of the actual market price.

To this point, the filtering parameter, $C$, has been constrained to
1, thus preserving coefficients that are less than the average
coefficient at a particular scale.  However, one might wonder if it is
possible to tune this parameter to maximally remove the uninteresting
Gaussian noise from the original signal.

\begin{figure}
\centerline{\epsfig{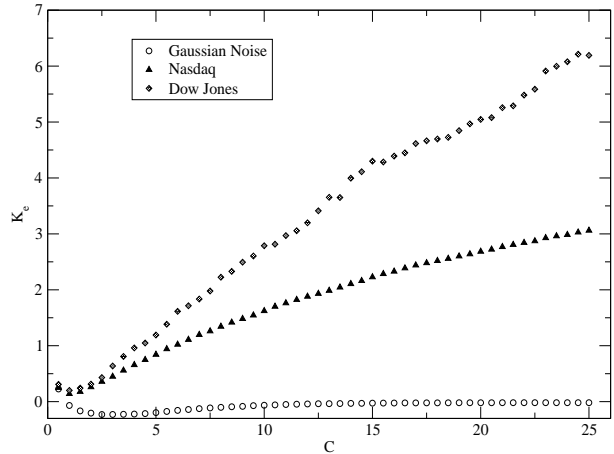}}
\caption{ The excess of kurtosis plotted as a function of the filter
  parameter $C$ for the NQ and DJ indices.  A sample of Gaussian noise
  is also included for contrast.  An optimal value of $C \sim 1$ is
  found, optimally filtering the original market time series to the
  level of noise.}
\label{kur_pl}
\end{figure}

Fig.~\ref{kur_pl} illustrates the extent to which the filtered signal
is Gaussian as a function of the filtering parameter $C$.  Here we
report the value of the excess of kurtosis, $K_e= \langle r^4 \rangle
/\langle r^2 \rangle ^{2}-3$, where $\langle ... \rangle$ is the
average of the filtered time series over the period considered.  For 
pure Gaussian noise this value should be 0.  With this test
we are able to identify $C \sim 1$ as optimal for both the NQ and DJ
indices investigated here.  Pure noise signals are
completely filtered with this simple consideration:
an examination of the standard autocorrelation function of the
filtered time series shows a complete temporal independence, further
confirming that we have successfully filtered Gaussian noise.

\section{Data Analysis}
\label{data_analysis}

Once we have isolated and removed noise from the time series we are
able to perform a reliable statistical analysis on the avalanches of
the residual returns, Fig.~\ref{filt_pl} (Bottom).  In particular, we
{\em define} an {\em avalanche} as the periods of the residual returns
in which the volatility, $v(t)\equiv |r(t)|$, is above a small
threshold, typically two orders of magnitude smaller than the
characteristic return.

A parallel between avalanches in the classical sandpile models (BTW
models) exhibiting SOC~\cite{Bak8788} and the previously defined
coherent events in the stock market is straightforward.  In order to
test the relation between the two, we make use of some properties of
the BTW models.  In particular, we use the fact that the avalanche
size distribution and the avalanche duration are distributed according
to power laws, while the laminar, or waiting times between avalanches
are exponentially distributed, reflecting the lack of any temporal
correlation between them~\cite{Boffetta99,Wheatland98}.  This is
equivalent to stating that the triggering process has no memory.

Similar to the dissipated energy in a turbulent flow, we define an
avalanche size, $V$, 
in the market context 
as the integrated value of the squared volatility over each coherent
event of the residual returns.  The duration, $D$, is defined as the
interval of time between the beginning and the end of a coherent
event, while the laminar time, $L$, is the time elapsing between the
end of an event and the beginning of the next.

\begin{figure}
\centerline{\epsfig{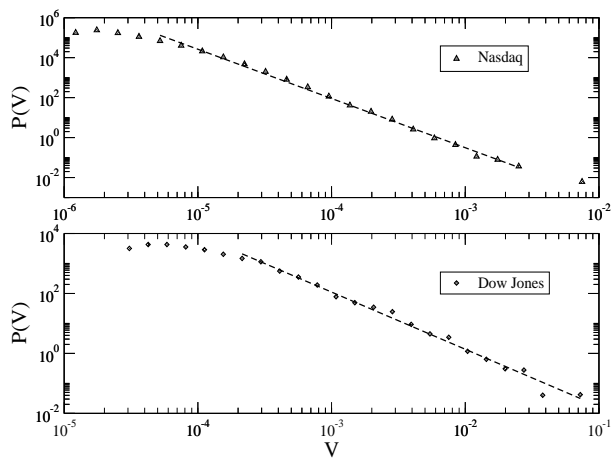}}
\caption{ PDFs for the avalanche size, $V$, for the optimally-filtered
  NQ (Top) and DJ (Bottom) indices.  The exponents of the power laws
  are $\gamma \sim -2.4$ (NQ) and $\gamma \sim -1.9$ (DJ).}
\label{ene_pl}
\end{figure}

\begin{figure}
\vspace{0.4cm}
\centerline{\epsfig{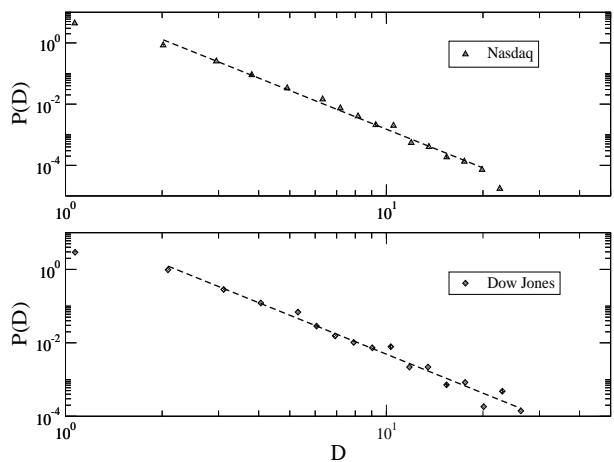}}
\caption{PDFs for the avalanche duration, $D$, for the
  optimally-filtered NQ (Top) and DJ (Bottom) indices.  The exponents
  of the power laws are $\gamma \sim -4.2$ (NQ) and $\gamma \sim -3.5$
  (DJ). }
\label{dur_pl}
\end{figure}

\begin{figure}
\vspace{0.85cm}
\centerline{\epsfig{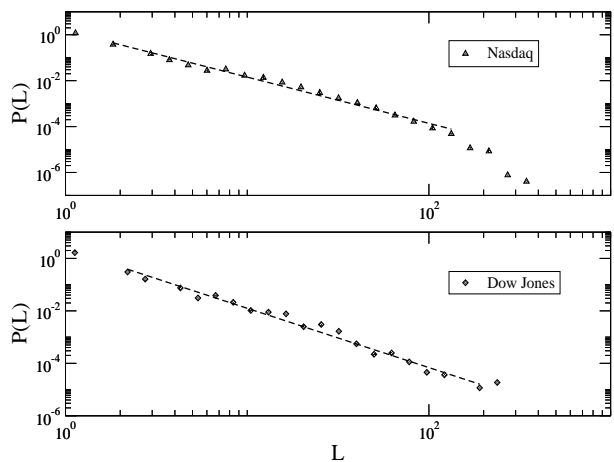}}
\caption{PDFs for the laminar times between avalanches, $L$, for the
  optimally-filtered NQ (Top) and DJ (Bottom) indices. The exponents
  of the power laws are $\gamma \sim -2.1$ (NQ) and $\gamma \sim -2.3$
  (DJ). }
\label{lam_pl}
\end{figure}

The results for the statistical analysis of the optimally-filtered NQ
and DJ indices are shown in Figs.~\ref{ene_pl}, \ref{dur_pl} and
\ref{lam_pl} for the avalanche size, duration and laminar times,
respectively.  A power law relation is clearly evident for all three
quantities investigated.

The data analyzed herein display a distribution of laminar times
different from the BTW model of the classical sandpile.  As explained
previously, the BTW model shows an exponential distribution for $L$,
derived from a Poisson process with no
memory~\cite{Boffetta99,Wheatland98}.  The power law distribution
found here implies the existence of temporal correlations between
coherent events.  However this correlation may have its origin in the
driver of the market, contrasting the random driver of the classical
sandpile.

%

\section{Discussion and Conclusion}

We have investigated the possible relations between the theory of
self-organized criticality and the stock market. The existence of a
SOC state for the latter would be of great theoretical importance, as
this would impose constraints on the dynamics, as implied by the
presence of a bounded attractor in the state space.  Moreover, it
would be possible to build new predictive schemes based on this
framework~\cite{Caglioti95}.

After a multiscale wavelet filtering, an avalanche-like dynamics has
been revealed in two samples of market data.
The avalanches are characterized by a scale-free behaviour in the
size, duration and laminar times.  The power laws in the avalanche
size and duration are a characteristic feature of a critical
underlying dynamics in the system.

However, the power law behavior in the laminar time distribution
implies a memory process in the triggering driver that is absent in
the classical BTW models, where an exponential behavior is expected.
Remarkably, the same features have been also observed in other
physical
contexts~\cite{Boffetta99,Kovacs01,Spada01,Antoni01,Corral04}.

The problem of temporal correlation in the avalanches of real systems,
has raised debates in the physics community, questioning the practical
applicability of the SOC framework~\cite{Carbone02}.  Motived by this
issue, several numerical studies have been devoted to including
temporal correlations in SOC
models~\cite{Norman01,Lippiniello05,Baiesi05}.  A power-law
distribution in the laminar times has been achieved, for example, by
substituting the random driver with a chaotic
one~\cite{DeLosRios97,Sanchez02}.  Alternatively, it has been shown
that non-conservative systems, as for the case of the stock market,
could be in a {\em near-SOC} state where dissipation induces temporal
correlations in the avalanches while the power law dynamics persist
for the size and duration~\cite{Freeman00,Carvalho00}.

In conclusion, a definitive relation between SOC theory and the stock
market has not been found.  Rather, we have shown that a memory
process is related with periods of high activity.  The memory could
result from some kind of dissipation of information, similar to
turbulence, or have its origin in a chaotic driver applied to the
self-organized critical system.  While a combination of the two
processes can also be possible, it is the latter property that
prevents one from ruling out the possibility that the stock market is
indeed in a SOC state \cite{Sanchez02}.

Similar power-law behaviour has been found in the ASX index for the
Australian market \cite{SOC1} and different single stock time
series.  
If this power-law behaviour is confirmed by further studies, this
should be considered as a stylized fact of stock market dynamics.


\section*{Acknowledgements}
This work was supported by the Australian Research Council.


\begin{thebibliography}{00}
\bibitem{mantegna}
R. N. Mantegna and H. E. Stanley, \textit{An Introduction to Econophysics: 
Correlation and Complexity in Finance}, (Cambridge University Press, Cambridge, 1999).
\bibitem{Bak93b}P. Bak  {\em et al.}, Ric. Econ. {\bf 47}, 3 (1993).
\bibitem{Bak97}P. Bak  {\em et al.}, Physica A {\bf 246}, 430 (1997).
\bibitem{Turcotte99}D. L. Turcotte, Rep. Prog. Phys. {\bf 62}, 
1377 (1999).
\bibitem{Feigenbaum03} J. Feigenbaum, Rep. Prog. Phys. {\bf 66}, 1611 (2003).
\bibitem{Bartolozzi05} M. Bartolozzi, D.B. Leinweber and A.W. Thomas, Physica A
{\bf 365}, 449 (2006).
\bibitem{Bak8788} P. Bak  {\em et al.}, Phys. Rev. Lett. {\bf 59}, 381 (1987);
P. Bak  {\em et al.}, Phys. Rev. A {\bf 38}, 364 (1988).
\bibitem{Jensen}H. J. Jensen, \textit{Self-Organized Criticality: 
Emergent Complex Behavior in Physical and Biological Systems},
(Cambridge University Press, Cambridge, 1998).
\bibitem{SOC1} M.~Bartolozzi, D.~B.~Leinweber and A.~W.~Thomas, 
Physica A {\bf 350}, 451 (2005).
\bibitem{Caglioti95} E. Caglioti and V. Loreto, Phys. Rev. E, {\bf 53}, 2953 (1996).
\bibitem{Frisch} U. Frisch, \textit{Turbulence}, (Cambridge University Press, Cambridge, 1995).
\bibitem{Ghashghaie96} S. Ghashghaie {\em et al.}, Nature {\bf 381}, 
767 (1996).
\bibitem{Mantegna97} R. N. Mantegna and H. E. Stanley, Physica A {\bf 239}, 
225 (1997).
\bibitem{Farge92} M. Farge, Annu. Rev. Fluid Mech. {\bf 24}, 395 (1992).
\bibitem{Farge99} M. Farge {\em et al.}, Phys. Fluids {\bf 11}, 2187 (1999).
\bibitem{Daubechies88} I. Daubechies, Comm. Pure Appl. Math. {\bf 41} (7), 909 (1988).
\bibitem{Kovacs01} P. Kov${\rm\acute{a}}$cs {\em et al.} Planet. Space Sci. {\bf 49}, 1219 (2001).
\bibitem{Katul94} G.G. Katul {\em et al.}, \textit{Wavelets in Geophysics}, 
pp. 81-105, (Academic, San Diego, Calif. 1994)
\bibitem{Boffetta99} G. Boffetta {\em et al.}, Phys. Rev. Lett. {\bf 83}, 4662 (1999).
\bibitem{Wheatland98} M.S. Wheatland {\em et al.}, Astrophys.J.{\bf 509}, 448 (1998).
\bibitem{Spada01} E. Spada {\em et al.}, Phys. Rev. Lett. {\bf 86}, 3032 (2001).
\bibitem{Antoni01} V. Antoni  {\em et al.}, Phys. Rev. Lett. {\bf 87}, 045001 (2001).
\bibitem{Corral04} A. Corral, Phys. Rev. Lett. {\bf 92}, 108501 (2004).
\bibitem{Carbone02} V. Carbone {\em et al.}, Europhys. Lett.,{\bf 58} (3), 
349 (2002).
\bibitem{Norman01} J.P. Norman {\em et al.}, Astrophys. J.,
 {\bf 557}, 891 (2001).
\bibitem{Lippiniello05} E. Lippiniello L. de Arcangelis and C. Godano,
Europhys. Lett.,{\bf 72}, 678 (2005).
\bibitem{Baiesi05} M. Baiesi and C. Maes, preprint: cond-mat/0505274.
\bibitem{DeLosRios97} P. De Los Rios {\em et al.}, Phys. Rev. E {\bf 56}, 4876 (1997).
\bibitem{Sanchez02} R. Sanchez {\em et al.}, Phys. Rev. Lett. 
{\bf 88}, 068302-1 (2002).
\bibitem{Freeman00} M.P. Freeman {\em et al.}, Phys. Rev. E {\bf 62}, 8794 (2000).
\bibitem{Carvalho00} J.X. Carvalho and C.P.C. Prado, Phys. Rev. Lett.,
 {\bf 84}, 4006 (2000).

\end{thebibliography}
\end{document}